# Detection of multipolar orders in the spin–orbit-entangled $5d$ Mott insulator $Ba_2MgReO_6$

Daigorou Hirai[1]*, Hajime Sagayama[2], Shang Gao[3]†, Hiroyuki Ohsumi[4], Gang Chen[5,6], Taka-hisa Arima[3,7], and Zenji Hiroi[1]

[1] Institute for Solid State Physics, University of Tokyo, Kashiwa, Chiba 277-8581, Japan

[2] Institute of Materials Structure Science, High Energy Accelerator Research Organization, Tsukuba, Ibaraki 305-0801, Japan

[3] RIKEN Center for Emergent Matter Science, Wako 351-0198, Japan.

[4] RIKEN SPring-8 Center, Sayo 679-5148, Japan.

[5] Department of Physics and Center of Theoretical and Computational Physics, the University of Hong Kong, Hong Kong, China

[6] State Key Laboratory of Surface Physics and Department of Physics, Fudan University, Shanghai 200433, China

[7] Department of Advanced Materials Science, University of Tokyo, Kashiwa 277-8561, Japan

* E-mail: dhirai@issp.u-tokyo.ac.jp

† Present address: Materials Science & Technology Division and Neutron Science Division, Oak Ridge National Laboratory, Oak Ridge, TN 37831, USA

Abstract

**In electronic solids with strong spin–orbit interactions (SOIs), the spin and orbital degrees of freedom of an electron are quantum mechanically entangled, which may result in an exotic multipolar order instead of a conventional dipolar order such as a magnetic order. Such a higher-degree order is called "hidden order" because of difficulties in experimental detection. Moreover, the number of candidate compounds is limited, especially rare in *d* electron systems, in which an interplay between SOIs and Coulomb interactions is expected to cause rich physics. Here, we employ state-of-the-art synchrotron X-ray diffraction techniques on a high-quality single crystal to probe subtle symmetry breaking induced by a multipolar order. We unequivocally demonstrate that the double-perovskite $Ba_2MgReO_6$ exhibits successive transitions to quadrupolar and then dipolar orders upon cooling, which is consistent with a theory considering SOIs. Our findings are a significant step towards understanding the intriguing physics of multipoles realized by spin–orbit-entangled $5d$ electrons.**

Exotic quantum phases such as high-temperature superconductivity[1,2], colossal magnetoresistance[3,4], and heavy Fermion states[5,6] are observed in materials with strong electron–electron correlations, and both the spin and orbital degrees of freedom of an electron are thought to play a role. In particular, strong interactions between spin and orbital moments – an effect known as spin–orbit interactions (SOIs) – are expected to produce a variety of exotic phenomena[7–9].

In this context, over the past ten years, research has focused on heavy transition metal (TM) compounds[7–9], where the combined effect of electron correlations and SOIs is realized. For example, in $Sr_2IrO_4$, a spin–orbit-entangled Mott insulating state emerges as a result of electron correlation effectively enhanced by SOIs[10,11]. In addition, when these spin–orbit-entangled electrons interact with each other through specific direction-dependent interactions, a novel quantum phase called the Kitaev spin liquid occurs[9,12]. Moreover, exotic topological phases can emerge when the spin–orbit-entangled electrons begin to delocalize[7,8]. Despite these intriguing phenomena, the understanding of $5d$ electron systems remains incomplete. In particular, the most fundamental symmetry-breaking phase, that is, the multipolar order[13,14], has not yet been experimentally well established. The electrons with strong spin–orbital entanglement, denoted by a total angular momentum $J$, may experience various symmetry breaking transitions. These transitions result in complex angular distributions of spin and charge densities, which can be described by the quantum mechanically defined multipole moments[15]. The higher-order multipolar order is generally subtle and hard to detect by traditional experimental probes in comparison to conventional dipolar orders; this is why the multipolar order is also known as the "hidden order"[16,17]. In addition, model $5d$ materials that can demonstrate the characteristics of multipolar orders are lacking thus far.

There are two requirements to create a multipolar order in an actual material. One is a high local symmetry at the TM site, which leads to an unquenched orbital moment. The other is an appropriate distance and charge transfer between TM ions for electrons to be localized; many $5d$ TM compounds are weakly correlated metals with relatively large bandwidths. In terms of these requirements, double-perovskite (DP) compounds[18] are good candidates, as they comprise octahedrally-coordinated TM ions that are spatially separated from each other (Fig. 1a). Thus far, several DP compounds have been suggested by both theoretically[8,13,19] and experimentally to exhibit multipolar order: $Ba_2NaOsO_6$[20–22], $Ba_2MOsO_6$ (M = Zn, Mg, Ca)[23], $Ba_2YMoO_6$[24], $Ba_2MgReO_6$[25,26], and $A_2TaCl_6$ (A = Rb, and Cs)[27]. A recent nuclear magnetic resonance (NMR) study on $Ba_2NaOsO_6$ revealed that the compound shows a noncollinear spin order and a small structural distortion, which was suggested to be due to a quadrupolar order[22]. However, the order parameter or ordering structure was not identified. Moreover, ferro-octupolar order was suggested for $Ba_2MOsO_6$ (M = Zn, Mg, Ca) based on the observation of a gapped magnetic excitation spectrum without additional magnetic Bragg peaks[23].

In this article, we focus on $Ba_2MgReO_6$[25,26,28]. Using state-of-the-art synchrotron X-ray diffraction (XRD) techniques on high-quality single crystals, we have successfully observed orderings of both magnetic dipoles at the magnetic transition temperature $T_m$ = 18 K and charge quadrupoles at the quadrupolar order temperature $T_q$ = 33 K and unambiguously determined their patterns for the first time. Our observations are in good agreement with the theoretical model that considers strong SOIs[13]. This compound provides us with a new opportunity to study quantum phases appearing among strongly correlated electrons with strong SOIs.

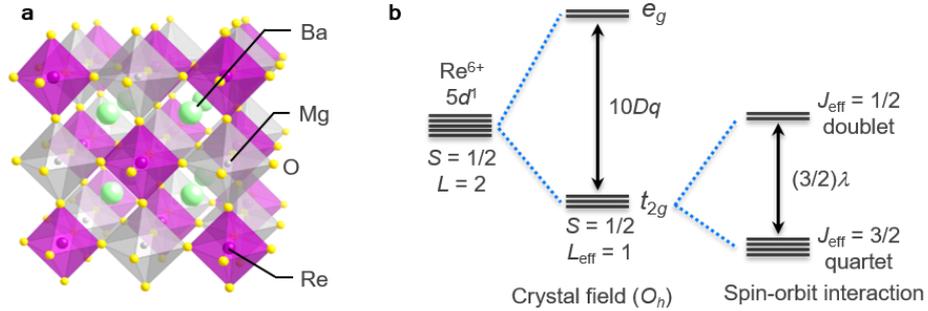

**Fig. 1 | Crystal structure and energy diagram for $Ba_2MgReO_6$.** **a** Cubic $Fm\text{–}3m$ structure of $Ba_2MgReO_6$ features a corner-sharing network composed of $ReO_6$ and non-magnetic $MgO_6$ octahedra arranged in a rock-salt structure. **b** Energy diagram of a single electron occupying the $5d$ orbitals in a $Re^{6+}$ ion. The crystal field from the surrounding six oxygen ions splits the fivefold-degenerate $5d$ states by $10Dq \sim 4$ eV to a triply degenerate $t_{2g}$ state and a doubly degenerate $e_g$ state. A strong spin–orbit interaction $\lambda$ as large as 0.5 eV further splits the lower-lying $t_{2g}$ state by $(3/2)\lambda$ into a $J_{eff}$ = 1/2 ($\Gamma_7$) doublet and a $J_{eff}$ = 3/2 ($\Gamma_8$) quartet by mixing the spin angular momentum ($S$ = 1/2) and effective orbital angular momentum ($L_{eff}$ = 1). The spin–orbit-entangled $J_{eff}$ = 3/2 quartet is the ground state of $Ba_2MgReO_6$ and has a multipolar degree of freedom.

$Ba_2MgReO_6$ is a $5d$ Mott insulator[25,26] that crystallizes in a DP (elpasolite) structure with a

face-centred cubic lattice in the space group *Fm–3m*, as illustrated in Fig. 1a. The $Re^{6+}$ ion possesses a $5d^1$ electron configuration and adopts a spin–orbit-entangled $J_{eff} = 3/2$ quartet as the ground state[29] (Fig. 1b). This $J_{eff} = 3/2$ quartet state has been inferred by the large reduction in the paramagnetic moment[26] (Fig. 2a): $0.68\mu_B$ instead of the expected value of $1.73\mu_B$ for $S = 1/2$. In addition, the electronic entropy released at low temperatures is 11.3 J $K^{-1}$ $mol^{-1}$, which is close to the electronic entropy value of $R\ln4 = 11.4$ J $K^{-1}$ $mol^{-1}$ (where $R$ is the gas constant) expected for a quartet state[26] (Fig. 2b).

Upon cooling, $Ba_2MgReO_6$ undergoes two phase transitions at $T_q = 33$ K and $T_m = 18$ K, as shown by the temperature dependences of the inverse magnetic susceptibility and heat capacity in Figs. 2a and 2b, respectively. In the magnetically ordered phase below $T_m$, an exotic spin structure is inferred from the small saturation moment of $0.3\mu_B$ and an unusual easy-axis along the [1 1 0] direction, the latter of which is incompatible with standard Landau theory for a cubic ferromagnet (Fig. 2c). These magnetic behaviours are very similar to those observed in $Ba_2NaOsO_6$[22]. Thus, a quadrupolar order is expected to occur at high temperatures above $T_m$, which may correspond to the intermediate phase between $T_q$ and $T_m$ in $Ba_2MgReO_6$. However, the nature of the intermediate phase remains unknown. In particular, a key evidence for the quadrupolar order, such as a structural change accompanied by the quadrupolar order, has not yet been detected in previous neutron diffraction measurements on polycrystalline samples[25].

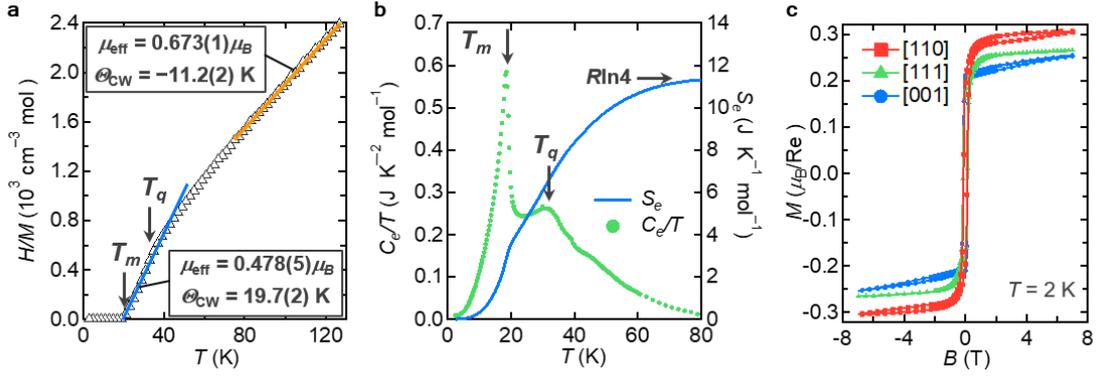

**Fig. 2 | Successive phase transitions in Ba$_2$MgReO$_6$[24]. a** Temperature dependence of the inverse magnetic susceptibility of a crystal in a magnetic field of 0.1 T applied in the [1 1 1] direction. Note that there are two Curie–Weiss regimes with different values of the effective magnetic moment, $\mu_{eff}$, and the Weiss temperature, $\Theta_{CW}$, across $T_q$ = 33 K: the solid blue and orange lines show Curie–Weiss fits of the data at 20–33 K and 80–350 K, respectively. **b** Temperature dependences of the electronic heat capacity divided by temperature ($C_e/T$, green circles, left scale), and the electronic entropy deduced from $C_e/T$ ($S_e$, blue line, right scale). The entropy released below 80 K is close to the value of $R\ln4$ = 11.5 J K$^{-1}$ mol$^{-1}$ expected for a quartet ground state. **c** Field dependences of magnetization of a crystal at 2 K in magnetic fields applied along the [1 1 0], [1 1 1], and [1 0 0] axes. There is marked anisotropy in the magnitude and field dependence near saturation, which is explained by the spin model depicted in Fig. 3c.

To reveal the characteristics of the successive phase transitions in Ba$_2$MgReO$_6$, we employed two kinds of XRD techniques using synchrotron radiation. First, the magnetic order below $T_m$ has been investigated in terms of the resonant effect at the X-ray absorption edge, which is generally superior in determining the ordered structures of small dipole moments using small single crystals[11,30–32]. Second, structural changes across the two transitions at $T_q$ and $T_m$ are examined by non-resonant XRD experiments to probe the quadrupolar order and its influence on the magnetic order. The very small structural change accompanied by the quadrupolar order, which is approximately one order of magnitude smaller than that induced by the orbital order in 3$d$ electron systems[33,34], can only be detected with ultra-high-intensity synchrotron X-ray beam using high-crystallinity samples.

Figure 3a shows a typical XRD profile taken below $T_m$ at 6 K, for the (1 0 0)$_t$ reflection which is not allowed for the body-centered-tetragonal lattice attained below $T_q$, as described later; the subscripts 't' and 'c' denote the index based on the tetragonal and cubic cells, respectively. The peak is strongly enhanced at around the Re $L_{III}$ absorption edge of 10.535 keV, indicative of its magnetic origin. Moreover, polarization analysis (shown in Supplementary Fig. S1) shows that the reflection appears not in the σ–σ′ channel but in the σ–π′ channel, further confirming it is a magnetic reflection. This (1 0 0)$_t$ magnetic reflection appears only below 18 K (Fig. 3b), indicating that the dipole moments of Re arranged below $T_m$.

We observe four magnetic reflections exhibiting resonance enhancement in our resonant XRD experiments (shown in Supplementary Fig. S2), all of which can be indexed with a single propagation vector with $\boldsymbol{k}$ = [0 0 1]$_t$. This propagation vector indicates a two-sublattice magnetic structure with the spins aligning ferromagnetically within the (0 0 1)$_t$ plane and with the layers stacking antiferromagnetically along the [0 0 1]$_t$ direction. Thus, the observed net moments must be parasitic, arising from spin canting. Considering the easy-axis anisotropy along the [1 1 0]$_c$ direction in the cubic structure[26], we propose a possible spin structure as depicted in Fig. 3c: the spins in the $z = 0$ plane rotate from [1 1 0]$_c$, which corresponds to [1 0 0]$_t$, by an angle of $+\phi$ and those in the $z = 1/2$ plane rotate by $-\phi$, so that an uncompensated moment remains along [1 1 0]$_c$. This canted antiferromagnetic spin structure is essentially the same as that theoretically predicted for the spin–orbit-entangled $d^1$ electron in DPs[13,35] and observed for Ba$_2$NaOsO$_6$[22]. The canting angle is large: $\phi$ ~ 40° for Ba$_2$MgReO$_6$[26] and 67° for Ba$_2$NaOsO$_6$[22]. This origin of unusual magnetic order may be related to the theoretically predicted quadrupolar order above $T_m$.

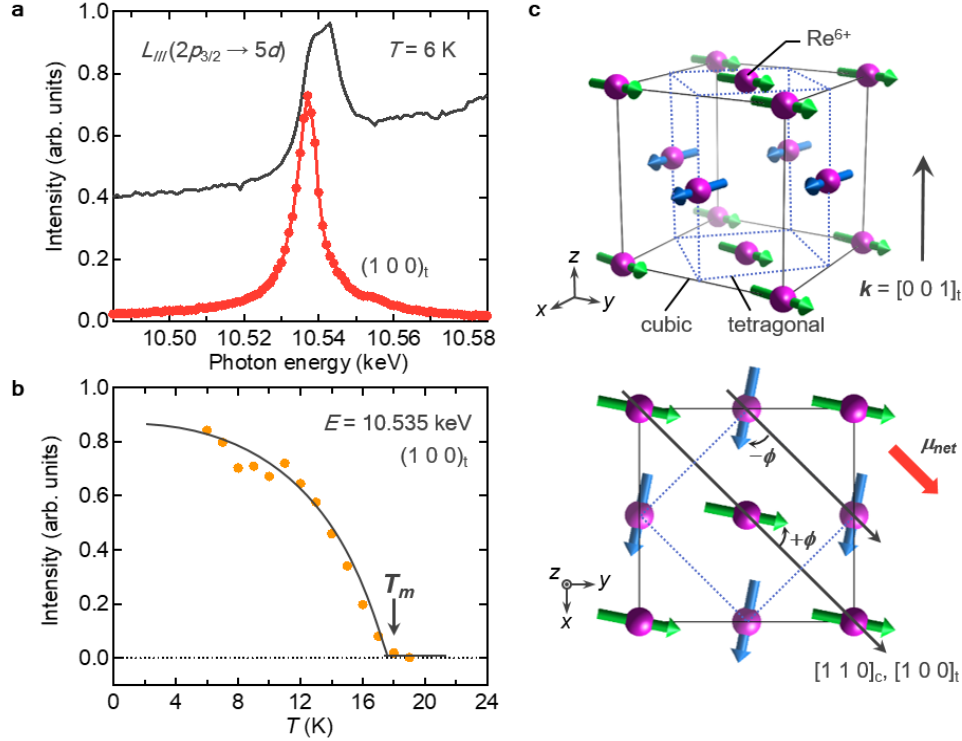

**Fig. 3 | Magnetic dipole transition and order in Ba$_2$MgReO$_6$. a** Energy dependence of the (1 0 0)$_t$ magnetic reflection intensity (red circles) and the X-ray fluorescence spectrum (solid line) measured below $T_m$ at 6 K. The subscripts 't' and 'c' denote the index based on the tetragonal and cubic cells shown in **c**. The (1 0 0)$_t$ peak intensity exhibits a large resonant enhancement at the Re $L_{III}$ absorption edge with an energy of approximately 10.535 keV, which corresponds to an excitation from the $2p_{3/2}$ state to the $5d$ state. **b** Temperature dependence of the intensity of the (1 0 0)$_t$ reflection (orange circles), which increases below $T_m = 18$ K. The black line is a visual guide. **c** Proposed magnetic structure with a propagation vector of $k = [0\ 0\ 1]_t$ as deduced from the resonant scattering and magnetization measurements. Magnetic moments on the two sublattices are distinguished by different colors; green at $z = 0$ and blue at $z = 1/2$. These moments rotate by angles of $+\phi$ and $-\phi$ from the cubic $[1\ 1\ 0]_c$ direction corresponding to the $[1\ 0\ 0]_t$ direction in the tetragonal cell, respectively, which results in a net moment $\mu_{net}$ in the $[1\ 1\ 0]_c$ direction. The high-temperature face-centered cubic and low-temperature primitive tetragonal unit cells are depicted by solid and dotted lines, respectively.

Next, we investigated the nature of the phase transition at $T_q = 33$ K using the non-resonant synchrotron XRD technique. As summarized in Fig. 4 and Supplementary D, a clear cubic-to-tetragonal structural transition is observed at $T_q$, which is evidenced by a split of the (0 0 24)$_c$ Bragg peak (inset of Fig. 4b). The relation between the high-temperature cubic and low-temperature tetragonal unit cells is depicted in Figure 3c. As shown in Fig. 4b, the tetragonal distortion ($c_t - \sqrt{2}a_t$) develops gradually below $T_q$, leading to a very small distortion of ~0.09% at 25 K (Supplementary Table S2). The distortion in $Ba_2MgReO_6$ is much smaller than in other DPs, for example, ~0.47% in $Ba_2CaWO_6$[36].

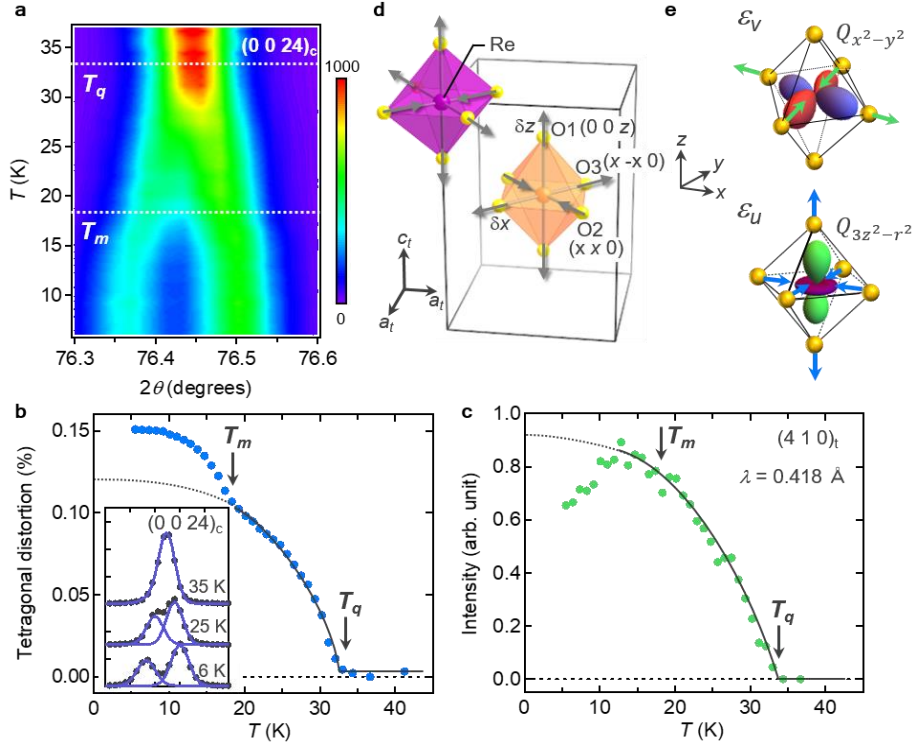

**Fig. 4 | Structural transition accompanied by a quadrupolar order in $Ba_2MgReO_6$.** **a** Temperature evolution of the synchrotron XRD intensity around the $(0\ 0\ 24)_c$ reflection measured on a single crystal of $Ba_2MgReO_6$. The peak splitting begins below $T_q$ and becomes more evident below $T_m$. **b** Temperature dependence of the tetragonal distortion, $c_t - \sqrt{2}a_t$, obtained by fitting the $(0\ 0\ 24)_c$ reflection with double Gaussian functions, as shown for the typical 6, 25, and 35 K data in the inset. Note that the tetragonal distortion is further enhanced upon cooling below $T_m$. **c** Temperature dependence of the integrated intensity of the $(4\ 1\ 0)_t$ reflection. The black lines in **b** and **c** are visual guides. **d** Schematic representations of the structural deformation observed below $T_q$. The $ReO_6$ octahedra at $z = 0$ and 1/2 are shown by purple and orange colors, respectively. The single type of oxygen site in the high-temperature cubic $Fm\text{-}3m$ structure separates into three types of sites in the low-temperature tetragonal $P4_2/mnm$ structure: O1, O2, and O3. The grey arrows show displacements of the oxygen atoms. **e** Schematic of the quadrupolar moments, $Q_{x2-y2}$ and $Q_{3z2-r2}$, which couple with the $\varepsilon_v$ and $\varepsilon_u$ modes observed in $Ba_2MgReO_6$. The green and blue arrows represent the atomic shifts in the $\varepsilon_v$ and $\varepsilon_u$ modes, respectively.

In addition to the split of fundamental reflections, 141 additional (superlattice) reflections, which are extremely weak (less than 0.005% of the strongest fundamental reflection), are detected below $T_q$ in our non-resonant X-ray diffraction experiments. The evolution of intensity over temperature of a superlattice reflection, $(4\ 1\ 0)_t$, is shown in Fig. 4c, appearing at $T_q$ and increasing gradually with increasing tetragonal distortion. Then, the intensity starts to decrease at approximately $T_m$, which is in contrast to the increase in the tetragonal distortion below $T_m$. We observe no additional superlattice reflections below $T_m$, suggesting no additional changes in the crystal structure. Extinction condition of $0\ 0\ l$: $l = 2n$, $h\ 0\ 0$: $h = 2n$, and $0\ k\ l$: $k + l = 2n$ are observed for the superlattice reflections (Supplementary *C*), suggesting the tetragonal space groups that are possible, namely, *P*4$_2$/*mnm* (136), *P*–4*n*2 (118), and *P*4$_2$*nm* (102). Note that none of these space groups are observed for other DP compounds[18], indicating that the structural distortion below $T_q$ is unique. The observed small distortion and distinguished crystal symmetry suggest that the structural change of Ba$_2$MgReO$_6$ is induced by a quadrupolar order.

Starting from a high-symmetry tetragonal space group *I*4/*mmm*, we consider atomic displacements in the intermediate phase (details are described in Supplementary *C* and *E*). Among the possible space groups, only *P*4$_2$/*mnm* belongs to the maximal non-isomorphic subgroups of *I*4/*mmm*; the other possible space groups of *P*–4*n*2 and *P*4$_2$*nm* are the maximal non-isomorphic subgroups of *P*4$_2$/*mnm*[37]. Our refinement reveals that space group *P*4$_2$/*mnm* provides a satisfactory fit, while a further symmetry lowering in the *P*–4*n*2 and *P*4$_2$*nm* space group does not improve the fit. As shown in Supplementary Fig. S4, this model accurately reproduces the intensities of all the observed (141) superlattice reflections as well as those of the fundamental reflections. In the tetragonal *P*4$_2$/*mnm* structure, from the high-temperature cubic *Fm*–3*m* structure, a single type of oxygen site splits into three types of oxygen site, as illustrated in Fig. 4d: an apical site O1 at 4*e* and in-plane sites O2 at 4*f* and O3 at 4*g*. This site splitting causes a slight elongation of the ReO$_6$ octahedron in the *c* direction and a rhomboid $\varepsilon_v$-type distortion of the square formed by four oxygens surrounding the Re ion in the (0 0 1) plane. The displacement of O1 from the regular octahedron ($\delta z$ at 6 K ~0.007 Å) corresponds to approximately half of the tetragonal distortion; $(c_t - \sqrt{2}a_t)/2$. The positions of O2 and O3 are described as $(x_{av}+\delta x\ x_{av}+\delta x\ 0)$ and $(x_{av}+\delta x\ -x_{av}-\delta x\ 0)$, respectively, where $x_{av}$ is the average *x* value and $\delta x$ is the in-plane shift (shown in Supplementary Fig. S3). The intensity of the superlattice peaks is approximately proportional to the square of $\delta x$. A comparison between the simulated and observed intensities provides an approximate estimate for $\delta x$ at 6 K of 0.022 Å.

The distortion of an octahedron, shown in Fig. 4d, is described by a combination of two normal modes of an octahedron, $\varepsilon_u$ and $\varepsilon_v$, which are depicted in Fig. 4e and Supplementary Fig. S5. Among the dipole, quadrupole, and octupole moments originating from the $J_{\text{eff}} = 3/2$ quartet[13], only a quadrupole moment, which is an anisotropic distribution of electronic charge, can induce lattice distortion by a linear coupling through electron–phonon interactions. The quadrupole moments in an octahedral field are classified into two-dimensional $\Gamma_3$ ($Q_{x2-y2}$, $Q_{3z2-r2}$) and three-dimensional $\Gamma_5$ ($Q_{xy}$, $Q_{yz}$, $Q_{zx}$) representations, which are analogous to the $e_g$ and $t_{2g}$ orbitals of the $3d$ electron, respectively. Among these moments, only the former quadrupolar moments are compatible with the observed lattice distortions: in other words, only $Q_{x2-y2}$ and $Q_{3z2-r2}$ can linearly couple with $\varepsilon_v$ and $\varepsilon_u$, respectively (Fig. 4e). The fact that the two distortions take place simultaneously in one octahedron means that the quadrupole moment is a linear combination of the $Q_{x2-y2}$ and $Q_{3z2-r2}$ moments. The $Q_{x2-y2}$ component may be predominant because the amplitude of the $\varepsilon_v$ mode is four times larger than that of the $\varepsilon_u$ mode at 6 K: 0.044(5) Å and 0.01(1) Å, respectively (Supplementary Fig. S5). Note that the $\varepsilon_v$ distortion is uniform in a layer and stacks in a staggered manner along the $c$ axis, while the $\varepsilon_u$ distortion is common for all the ReO$_6$ octahedra (Fig. 4d). This result indicates an antiferroic alignment of the $Q_{x2-y2}$ moment and a ferroic alignment of the $Q_{3z2-r2}$ moment, as depicted in Fig. 5. Thus, we conclude that below $T_q$, quadrupolar order composed of antiferroically arranged $Q_{x2-y2}$ moments and ferroically arranged $Q_{3z2-r2}$ is created in Ba$_2$MgReO$_6$.

According to the previous theoretical calculations on the electronic orders of the $5d^1$ DPs[13], three types of interactions between the spin–orbit-entangled electrons are considered: the nearest-neighbour antiferromagnetic exchange $J$, the nearest-neighbour ferromagnetic exchange $J'$, and the electric quadrupolar interaction $V$. Canted antiferromagnetic magnetic order with a large ferromagnetic moment along [1 1 0]$_c$ observed in Ba$_2$MgReO$_6$[26] and Ba$_2$NaOsO$_6$[22] is theoretically predicted to be the ground state for relatively large values of $V/J$ and $J'/J$. In addition, it has been predicted that a quadrupolar order spreads at higher temperatures above this magnetic order, as shown in the phase diagram (Fig. 5)[8,13]. Thus, starting from a disordered state of the quartet, successive transitions occur upon cooling first by removing the quadrupolar degree of freedom and then by removing the dipolar degree of freedom; these transitions are of second order. Our observations for Ba$_2$MgReO$_6$ are consistent with the theoretical predictions.

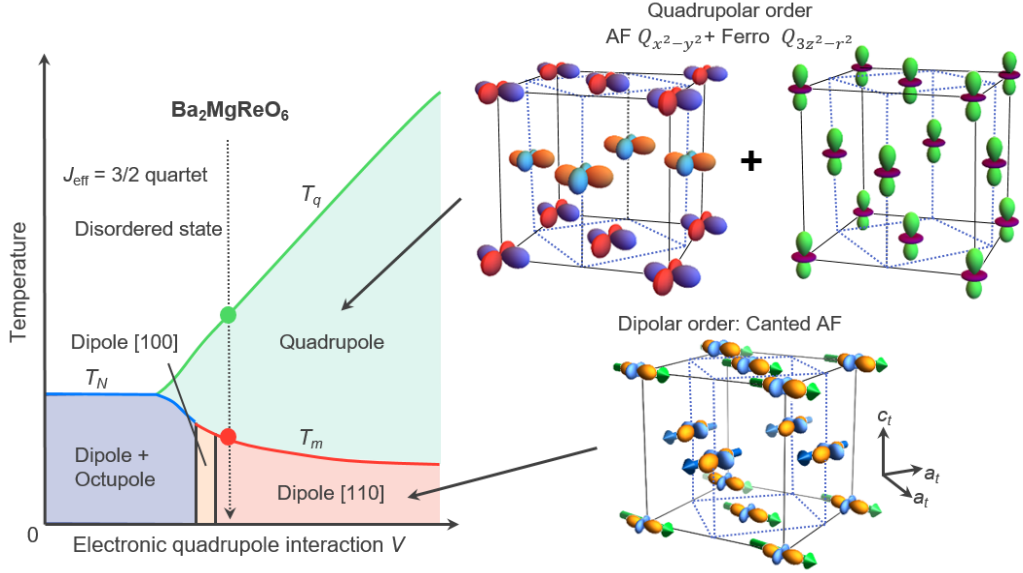

**Fig. 5 | Phase diagram and multipolar orders in DP Ba₂MgReO₆.** Schematic mean-field phase diagram for a $d^1$ double perovskite with a $J_{\text{eff}} = 3/2$ quartet state at $J'/J = 0.2$ is depicted as a function of the electric quadrupole interaction $V$[8,13]. 'Dipole [1 1 0]' and 'Dipole [1 0 0]' refer to the two-sublattice magnetic dipolar orders with the net magnetic moments along the $[1\ 1\ 0]_c$ and $[1\ 0\ 0]_c$ axes, respectively. 'Quadrupole' and 'Dipole + Octupole' refer to multipole orders of quadrupoles and of both dipoles and octupoles, respectively. The quadrupolar order below $T_q$ consists of a linear combination of a ferroic arrangement of $Q_{3z^2-r^2}$ quadrupolar moments and an antiferroic arrangement of $Q_{x^2-y^2}$ quadrupolar moments. The dipole order below $T_m$ is a canted antiferromagnetic order with the propagation vector $\mathbf{k} = [0\ 0\ 1]_t$, in which dipole moments are induced by modifying the $Q_{3z^2-r^2}/Q_{x^2-y^2}$ ratio of the preceding quadrupole order.

Anomalies in the temperature dependences of the tetragonality and the superlattice intensity at $T_m$, shown in Figs. 4b and 4c, are reasonably explained within the same theory. When the dipolar order occurs at $T_m$, other multipolar moments, such as octupolar and quadrupolar moments, are expected to be induced simultaneously[13]. The former does not cause lattice distortion, while a quadrupolar moment can induce the distortion of the ReO₆ octahedra. The experimental observations that the tetragonality is increased and the intensity of the $(4\ 1\ 0)_t$ superlattice reflection is decreased below $T_m$ indicate that the $Q_{3z^2-r^2}$ component is enhanced while the $Q_{x^2-y^2}$ component is reduced as the result of the dipolar order. Similar behaviour is observed for $3d$ TM compounds with Kugel-Khomskii couplings[38]; an antiferromagnetic order competes with a ferroic quadrupolar order and prefers an antiferroic one. Thus, antiferroically arranged $Q_{x^2-y^2}$ moments are decreased whereas ferroically arranged $Q_{3z^2-r^2}$ moments are increased with developing the antiferromagnetic order in $Ba_2MgReO_6$.

There is one notable discrepancy between our experimental results for $Ba_2MgReO_6$ and the theoretical prediction of the quadrupolar order[13]. We find a quadrupolar order with a linear combination of the $Q_{x^2-y^2}$ and $Q_{3z^2-r^2}$ moments below $T_q$ in $Ba_2MgReO_6$, whereas the theory predicts a simple antiferroic quadrupolar order of purely $Q_{x^2-y^2}$ moments, as depicted in the top left of Fig. 5. Notably, a similar type of quadrupolar (orbital) order with two components occurs in $LaMnO_3$, where electron–phonon couplings induce ferroic quadrupolar moments in the antiferroic quadrupolar order[39–41]. It is plausible that electron–phonon couplings, which are not considered in the theoretical calculation, and other effects such as quantum fluctuations play a role in stabilizing the $Q_{x^2-y^2}$–$Q_{3z^2-r^2}$ quadrupolar order in $Ba_2MgReO_6$.

In conclusion, our study establishes the existence of successive phase transitions to multipolar orders for the correlated spin–orbit–entangled 5$d$ electrons in the DP $Ba_2MgReO_6$. The quadrupolar order below $T_q$ = 33 K is composed of antiferroically arranged $Q_{x^2-y^2}$ and ferroically arranged $Q_{3z^2-r^2}$ moments. A noncollinear antiferroic arrangement of dipolar order appears at $T_m$ = 18 K, in which the quadrupolar order is significantly modulated. Our observation is consistent with the mean-field theory for spin–orbit–entangled electrons. Therefore, $Ba_2MgReO_6$ provides an opportunity to experimentally investigate the symmetry breaking of the multipolar degree of freedom in 5$d$ electron systems for the first time.

In future studies, the detection of elementary excitation from these multipolar orders must be of great interest and importance in capturing the dynamics of multipoles. There may be nontrivial excitations that are distinct from conventional magnetic excitations, as predicted by theories[42,43]. Also, it is highly desirable to construct a theory considering electron-phonon couplings or quantum fluctuations, which may confirm our experimental observations. Understanding the nature of interactions between multipolar moments would help us in exploring an exotic superconductivity mediated by multipolar fluctuations[44] and quantum liquid states of spin–orbit–entangled electrons[45,46].

**Acknowledgment**

The authors are grateful to R. Kurihara for insightful discussions. This work was partly supported by Japan Society for the Promotion of Science (JSPS) KAKENHI Grant Number JP18K13491, JP18H04308 (J-Physics) and by Core-to-Core Program (A) Advanced Research Networks. G.C. was supported by GRF No.17303819 from RGC of Hong Kong. The synchrotron radiation experiments at BL19LXU in SPring-8 were performed with the

**Method**

**Sample preparation** High-quality single crystals of $Ba_2MgReO_6$ were grown by the flux method in an enclosed space for controlling the oxidation state of rhenium ions. BaO, MgO, and $ReO_3$ powders with a ratio of 2:1:1 were mixed with a flux composed of $BaCl_2$ and $MgCl_2$ in an argon-filled glove box, and the mixture was sealed in a platinum tube. The tube was heated at 1300 °C and then slowly cooled to 900 °C, followed by furnace cooling to room temperature. Black shiny crystals with the octahedral morphology were obtained after the residual flux was washed away with distilled water.

**Resonant XRD** Resonant XRD experiments were performed on BL19LXU at SPring-8, Japan[47]. A $Ba_2MgReO_6$ single crystal with dimensions of $2 \times 2 \times 1$ mm$^3$ was fixed to a cold finger of a $^4$He closed-cycle refrigerator mounted on a four-circle diffractometer. The linearly polarized incident beam was referred to as the $\sigma$ polarization in a vertical scattering plane geometry and its intensity was monitored by an ionization chamber. The diffracted beam intensity was measured using a silicon drift detector (SDD). The azimuthal angle $\varphi$ was defined as $\varphi = 0°$ when the tetragonal $a$ axis was parallel to the X-ray polarization.

**Off-resonant XRD** Off-resonant XRD experiments were performed on the beamline BL-8A, Photon Factory (PF) and on the beamline AR-NE1A, Photon Factory Advanced Ring (PF-AR), in High Energy Accelerator Research Organization (KEK), Japan. Incident X-ray beams monochromatized by using Si (111) double crystals with wavelength $\lambda = 0.68987$ Å were used at BL-8A and high-flux and short-wavelength X-ray beams with $\lambda = 0.41827$ Å were used at AR-NE1A in order to detect weak superlattice reflections; these wavelengths were calibrated by measuring the diffraction profiles of a standard polycrystalline $CeO_2$ sample. A $Ba_2MgReO_6$ single crystal was attached to the top of a sapphire needle with Apiezon-N grease and mounted in a closed-cycle $^4$He refrigerator. XRD intensities were collected by the oscillation photograph technique using a large cylindrical image plate to index Bragg reflections and integrating their intensities. The RAPID-AUTO program developed by RIGAKU corp. was used on both the beamlines. Structural parameters were refined with the SHELXL program[48]. Real and imaginary parts of anomalous scattering factors calculated with D. T. Cromer and D. Liberman's method were used[49].

# *Supplementary Material:*

# Detection of multipolar orders in the spin–orbit-entangled 5*d* Mott insulator Ba$_2$MgReO$_6$

Daigorou Hirai, Hajime Sagayama, Shang Gao, Hiroyuki Ohsumi, Gang Chen, Taka-hisa Arima, and Zenji Hiroi

### A. *Polarization analysis for the (1 0 0)$_t$ magnetic reflection*

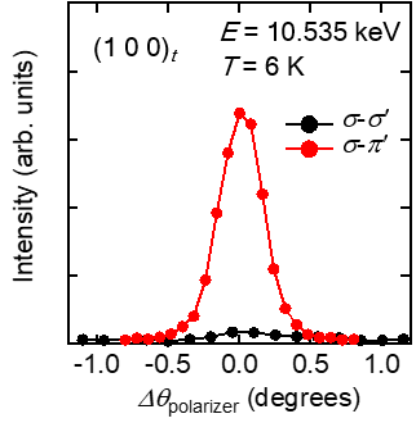

**Fig. S1 | Polarization analysis for the (1 0 0)$_t$ reflection of Ba$_2$MgReO$_6$.** The polarization state of the diffracted beam at the (1 0 0)$_t$ reflection was analyzed using the (0 0 8) reflection of pyrolytic graphite (PG) as a polarizer at the absorption edge of 10.535 keV at 6 K. Here $_t$ denotes tetroangoal $P4_2/mnm$ structure below the quadrupolar transition of 33 K. The rocking curve of the PG (0 0 8) reflection in the $\sigma$–$\sigma'$ and $\sigma$–$\pi'$ channels are shown by the black and red lines, respectively. The large intensity only in the $\sigma$–$\pi'$ channel evidences the magnetic origin of the (1 0 0)$_t$ reflection.

### B. *Magnetic reflections below $T_m$ at 6 K*

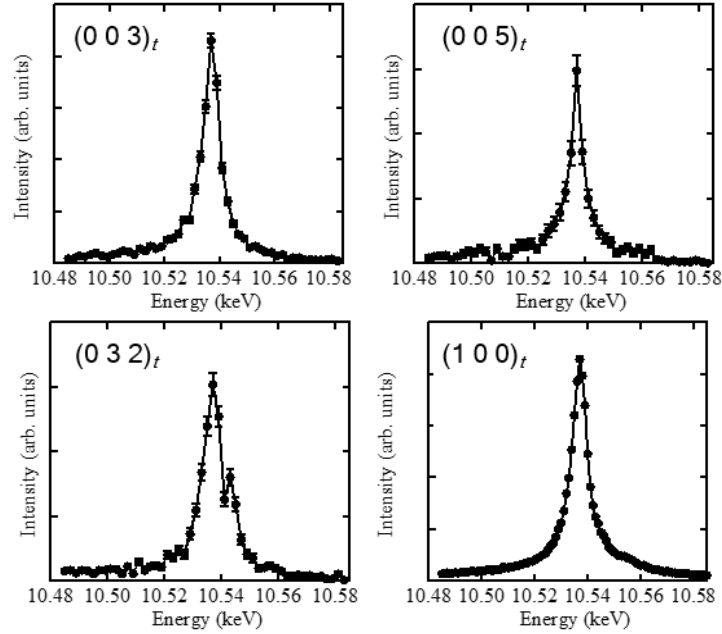

**Fig. S2 | Magnetic reflections observed below $T_m$ in Ba$_2$MgReO$_6$.** Energy dependences of the superlattice reflections observed below $T_m$ at 6 K. The peak intensities of observed four reflections exhibit a large resonance enhancement at the Re $L_{III}$ absorption edge with an energy of approximately 10.535 keV. The observed superlattice reflections (0 0 3)$_t$ and (0 0 5)$_t$, (1 0 0)$_t$, and (0 3 2)$_t$ violate reflection conditions for $P4_2/mnm$, 0 0 $l$: $l = 2n$, $h$ 0 0: $h = 2n$, and 0 $k$ $l$: $k + l = 2n$, respectively. All the indexes of these reflections are described by combinations of allowed reflections and the magnetic propagation vector $k$ = [0 0 1]$_t$: (0 0 2)$_t$ + $k$ = (0 0 3)$_t$, (0 0 4)$_t$ + $k$ = (0 0 5)$_t$, (1 0 –1)$_t$ + $k$ = (1 0 0)$_t$, and (0 3 1)$_t$ + $k$ = (0 3 2)$_t$.

## C. Possible space group for the quadrupolar order phase below $T_q$

We investigated the crystal structure of the quadrupolar order phase below $T_q$ by the non-resonant synchrotron X-ray experiments performed at the beamline AR-NE1A at Photon Factory Advanced Ring. Above $T_q$, a structural refinement at 40 K confirms an undistorted cubic structure of the space group *Fm–3m*, as reported previously (Supplementary Table S1). At $T_q$, we observe a clear structural change: several Bragg peaks split and 141 superlattice reflections appear. When we ignore the extremely weak superlattice reflections, the crystal structure below $T_q$ is reasonably refined by assuming a tetragonal space group *I4/mmm*, as shown in Supplementary Tables S2 and S3. Space group *I4/mmm* has the highest symmetry in the translationengleiche subgroups of the high-temperature cubic space group *Fm–3m*. Thus, the average structure has a space group *I4/mmm*.

Then, we consider further lowering of the symmetry from the average tetragonal *I4/mmm* structure. 141 superlattice reflections violate the extinction conditions for a body-centered tetragonal lattice. The observed superlattice reflections at (1 2 0), (1 1 1), and (1 2 2) violate reflection conditions $h\,k\,0$: $h + k = 2n$, $h\,h\,l$: $l = 2n$, $h\,k\,l$: $h + k + l = 2n$, respectively. On the other hand, since superlattice reflections at (0 0 3), (1 0 0), and (0 3 2) are not observed, there are extinction conditions: $0\,0\,l$: $l = 2n$, $h\,0\,0$: $h = 2n$, and $0\,k\,l$: $k + l = 2n$. These results provide a critical constraint to the tetragonal space groups that are possible, namely, *P4$_2$/mnm*, *P–4n2*, and *P4$_2$nm*. Since the $T_q$ transition is of the second order, we assume that *P4$_2$/mnm*, one of the maximal non-isomorphic subgroups of *I4/mmm*, is the actual space group; the other possible space groups are *P–4n2* and *P4$_2$nm* which are the maximal non-isomorphic subgroups of *P4$_2$/mnm*. As the atomic shifts estimated for *P4$_2$/mnm* are already very small, a further loss of symmetry (if it even exists) might be negligible and could be unimportant when considering the pattern of the quadrupolar order.

## D. Structural parameters for Ba$_2$MgReO$_6$

**Table S1 | Synchrotron X-ray single-crystal structure analysis of Ba$_2$MgReO$_6$ at 40 K in the paramagnetic phase.** $(x, y, z)$, $U_{iso}$, and $U_{nm}$ are the atomic coordinates, isotropic thermal parameter, and anisotropic thermal parameter, respectively. The temperature factor is expressed as $\exp(-2\pi^2 a^{*2}(U_{11}h^2 + U_{22}k^2 + U_{33}l^2))$. $U_{11} = U_{22} = U_{33} = U_{iso}$ for Re, Ba, and Mg atoms. The occupancy at each site is fixed to 1. The values in parenthesis represent standard deviations. The dataset for this analysis was collected at the beamline BL-8A at Photon Factory. The refinement was performed assuming space group of *Fm–3m*. 232 unique reflections were used for the refinement. Residual factors, $R_1$ (I > 2σ(I)) and $wR_2$ (all

reflections), and Goodness-of-fit indicator $S$ (all reflections), are 0.0124, 0.0299, and 1.151, respectively.

**Space Group $Fm$-$3m$, $a = b = c = 8.0802(2)$ Å, $α = β = γ = 90°$,**

|   | site | $x$ | $y$ | $z$ | $U_{iso}$ (Å$^2$) | $U_{11}$ (Å$^2$) | $U_{22}$ (Å$^2$) | $U_{33}$ (Å$^2$) |
|---|------|-----|-----|-----|-------------------|------------------|------------------|------------------|
| Re | 4b | 0.5 | 0 | 0 | 0.00073(4) | - | - | - |
| Ba | 8c | 0.25 | 0.25 | 0.25 | 0.00133(4) | - | - | - |
| Mg | 4a | 0 | 0 | 0 | 0.0030(3) | - | - | - |
| O | 24e | 0.26163(15) | 0 | 0 | - | 0.0030(3) | 0.0048(2) | 0.0048(2) |

**Table S2 | Synchrotron X-ray single-crystal structure analysis of Ba$_2$MgReO$_6$ at 25 K in the quadrupolar ordered tetragonal phase.** The dataset for this analysis was collected at the beamline BL-8A at Photon Factory. 568 unique reflections were used for the refinement. The refinement was performed assuming space group $I4/mmm$, for the average structure. $R_1$ (I > 2σ(I)), $wR_2$ (all reflections), and $S$ (all reflections), are 0.0203, 0.0445, and 1.101, respectively.

**Space Group $I4/mmm$, $a = b = 5.7119(5)$ Å, $c = 8.0849(7)$ Å, $α = β = γ = 90°$,**

|   | site | $x$ | $y$ | $z$ | $U_{iso}$ (Å$^2$) |
|---|------|-----|-----|-----|-------------------|
| Re | 2a | 0 | 0 | 0 | 0.00092(3) |
| Ba | 4d | 0 | 0.5 | 0.25 | 0.00152(4) |
| Mg | 2b | 0 | 0 | 0.5 | 0.0029(3) |
| O1 | 4e | 0 | 0 | 0.7616(2) | 0.00416(19) |
| O2 | 8h | 0.2383(2) | 0.2383(2) | 0 | 0.00437(14) |

**Table S3 | Synchrotron X-ray single-crystal structure analysis of Ba$_2$MgReO$_6$ at 6 K in the tetragonal magnetically ordered phase.** The dataset for this analysis was collected at the beamline AR-NE1A at Photon Factory Advanced Ring. 1424 unique reflections were used for the refinement. The refinement was performed assuming space group $I4/mmm$, for the average structure. $R_1$ (I > 2σ(I)), $wR_2$ (all reflections), and $S$ (all reflections), are 0.0309, 0.0972, and 1.232, respectively.

**Space Group $I4/mmm$, $a = b = 5.7107(2)$ Å, $c = 8.0893(2)$ Å, $α = β = γ = 90°$,**

|   | site | $x$ | $y$ | $z$ | $U_{iso}$ (Å$^2$) |
|---|------|-----|-----|-----|-------------------|
| Re | 2a | 0 | 0 | 0 | 0.00100(2) |
| Ba | 4d | 0 | 0.5 | 0.25 | 0.00146(3) |
| Mg | 2b | 0 | 0 | 0.5 | 0.0044(2) |
| O1 | 4e | 0 | 0 | 0.7611(7) | 0.0002(4) |
| O2 | 8h | 0.2384(9) | 0.2384(9) | 0 | 0.0084(6) |

### E. Estimation of the magnitude of oxygen displacements in the quadrupolar order phase

The magnitude of the displacements of oxygen atoms accompanied by the quadrupolar order is estimated based on the structural analysis using the off-resonant single-crystal X-ray diffraction data. The high-temperature $Fm\bar{3}m$ structure includes a single type of oxygen site, 24$e$ at ($x$ 0 0), and a single type of Re site, 4$a$ at (0 0 0), while the $P4_2/mnm$ structure includes three types of oxygen site, as illustrated in Fig. 4d: an apical site O1, 4$e$ at (0 0 $z$) and in-plane sites O2, 4$f$ at ($x$ $x$ 0) and O3, 4$g$ at ($x$ $-x$ 0). This site splitting causes a slight elongation of the ReO$_6$ octahedron in the $c$ direction and a rhomboid $\varepsilon_v$-type distortion of the square formed by four oxygens surrounding the Re ion in the (0 0 1) plane. The displacements of oxygen atoms in the tetragonal $P4_2/mnm$ structure for temperatures below $T_q$ is depicted in Fig. 4d.

Firstly, the elongation of the ReO$_6$ octahedron is estimated by performing structural refinements assuming the tetragonal space group $I4/mmm$, as the average structure. The result of structural refinements, as listed in Supplementary Table S3, yields two lengths of Re–O bonds at 6 K: 1.932(5) Å for the apical bond and 1.925(7) Å for the in-plane bond. Thus, a slight elongation of the ReO$_6$ octahedron by 0.4(4)% is induced in the $c$ direction. The estimated displacement of O1 from the regular octahedron corresponds to $\delta z$ of ~0.007 Å.

Secondly, the in-plane oxygen displacements are estimated by assuming the tetragonal space group $P4_2/mnm$. In the tetragonal $P4_2/mnm$ structure, the in-plane oxygen displacements are described by only one parameter $\delta x$ as shown in Fig S3. The structure factor of a superlattice reflection $F_{sl}(h\ k\ l)$ is derived as

$$F_{sl}(h\ k\ l) = 4f_O(\sin(2\pi(h+k)x_{av})\sin(2\pi(h+k)\delta x) - \sin(2\pi(h-k)x_{av})\sin(2\pi(h-k)\delta x)),$$
(S1)

where $f_O$ and $x_{av}$ are the atomic scattering factor of the divalent oxygen ion and the averaged position of the in-plane oxygen ions, respectively. Note that the superlattice reflections from the three directional domains ($c_t$ // [1 0 0]$_c$, [0 1 0]$_c$, and [0 0 1]$_c$) do not overlap each other because of the reflection condition. Thus, the volumes of the three domains are proportional to $F_{sl}(h\ k\ l)^2$ and are determined as 0.24, 0.37, and 0.39, respectively. Taking into account this volume ratio and comparing the observed intensities with the numerical simulations for both fundamental and superlattice reflections, we estimate $\delta x$ to be 0.00277(5) (~0.022 Å); the calculated $F_{sl}(h\ k\ l)^2$ reproduces very well the observed $F_{sl}(h\ k\ l)^2$ in a wide intensity range as shown in Fig. S4.

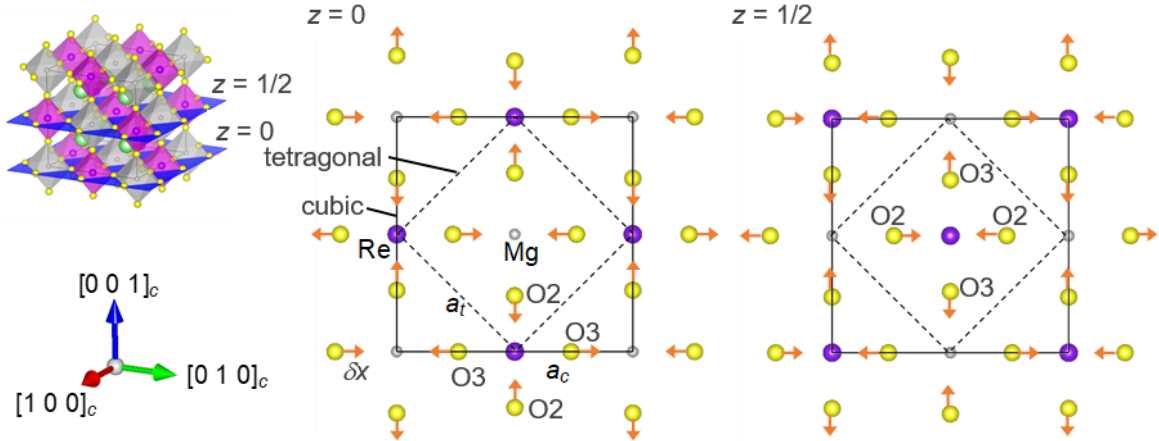

**Fig. S3 | In-plane Oxygen displacement pattern of $Ba_2MgReO_6$ in the low-temperature tetragonal $P4_2/mnm$ structure.** Two cross sections at $z = 0$ and $1/2$ are shown. The solid and broken squares represent the unit cells of the high-temperature cubic structure and the low-temperature tetragonal structure, respectively. The orange allows denote allowed atomic shifts of the in-plane oxygen atoms by $\delta x$.

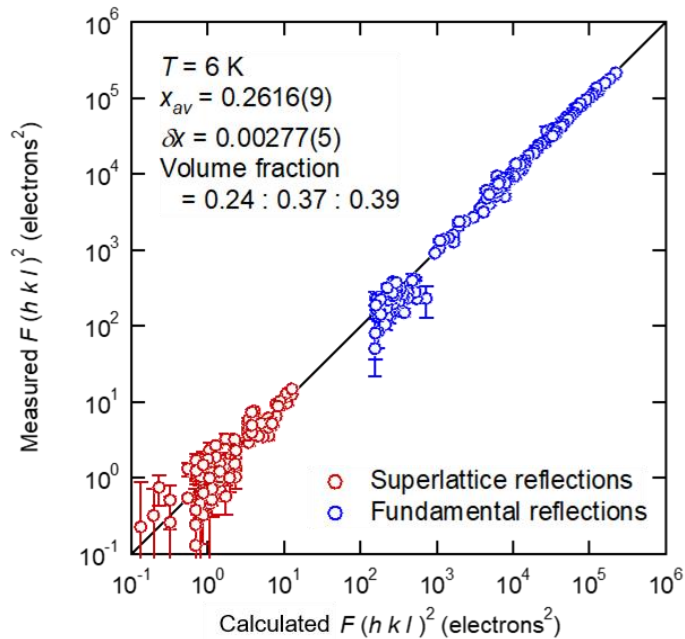

**Fig. S4 | Comparison of the observed $F(h\ k\ l)^2$ at 6 K and the calculated $F(h\ k\ l)^2$.** Blue circles and red circles denote the $F(h\ k\ l)^2$ of fundamental and 141 superlattice reflections, respectively.

## F. The normal modes of cubic lattice and its amplitude

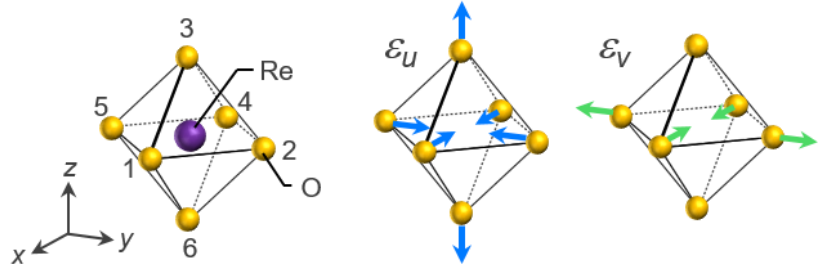

**Fig. S5 | Oxygen displacement patterns of the ReO$_6$ octahedron in Ba$_2$MgReO$_6$.** A regular oxygen octahedron in the high-temperature cubic phase (left) and distorted octahedra by the $\varepsilon_u$ mode (middle) and $\varepsilon_v$ mode (right) in the low-temperature tetragonal phase of the space group $P4_2/mnm$ are shown. The actual distortion is described by a linear combination of the two normal modes: $\varepsilon_u = (2z_3 - 2z_6 - x_1 + x_4 - y_2 + y_5)/2\sqrt{3}$ and $\varepsilon_v = (x_1 - x_4 - y_2 + y_5)/2$, where $x$, $y$ and $z$ are the coordinates of the oxygen atoms labelled 1–6. The blue and green arrows indicate the atomic shifts for the $\varepsilon_u$ and $\varepsilon_v$ modes, respectively. The amplitude of $\varepsilon_u$ and $\varepsilon_v$ modes are calculated based on the results of structural analysis assuming a space group $P4_2/mnm$, as described in Supplementary E. The obtained amplitudes of $\varepsilon_u$ and $\varepsilon_v$ modes at 6 K are 0.044(5) Å and 0.01(1) Å, respectively.